\documentclass[pra,twocolumn,showpacs,nobibnotes,floatfix,superscriptaddress]{revtex4}

\usepackage{latexsym,amsmath,amssymb,amsfonts,mathbbol,graphicx,color}
\usepackage{dcolumn}
\usepackage{bm}

\begin{document}

\title{Entanglement Sudden Death in a Quantum Memory}

\author{Yaakov S. Weinstein}
\affiliation{Quantum Information Science Group, {\sc Mitre},
260 Industrial Way West, Eatontown, NJ 07224, USA}


\begin{abstract}
I explore entanglement dynamics in examples of quantum memories, decoherence free 
subspaces (DFS) and noiseless subsystems (NS), to determine how a complete loss 
of entanglement affects the ability of these techniques to protect quantum 
information. Using negativity and concurrence as entanglement measures, I find that 
in general there is no correlation between the complete loss of entanglement in the 
system and the fidelity of the stored quantum information. These results complement 
previous results in which quantum protocols not explictly based on entanglement 
exhibit little correlation between ESD and the accuracy of the given protocol.
\end{abstract}

\pacs{03.67.Mn, 03.67.Bg, 03.67.Pp}

\maketitle

\section{Introduction}

Protecting quantum information from the affects of decoherence, unwanted 
interactions between the system and its environment, is a vital 
requirement of any hoped for quantum computer implementation \cite{book}.
Such protection can be achieved via active or passive techniques. 
Active techniques, quantum error correction, identify and correct errors
that may have affected the quantum information. Alternatively, passive 
techniques store the quantum information in such a way that it is
{\emph {a priori}} immune to error. Two general schemes which allow for the 
passive avoidance of quantum errors are decoherence free subspaces (DFS) 
and noiseless subsystems (NS). DFSs store quantum information in specific 
states with an inherent symmetry such that the information is then immune 
to decoherence generators that respect those symmetries \cite{ZR,DG}. 
NSs store quantum information in certain symmetries of other degrees
of freedom of the system \cite{KLV} which are not
the states themselves. The error avoidance properties of DFSs 
and NSs makes them especially well suited for the construction of quantum 
memories since the system storing the information can, ideally, never 
be addressed until the information is needed.  However, manifestations of both 
schemes may utilize states that are highly entangled and thus may 
be subject to entanglement sudden death (ESD). 

Entanglement is a uniquely quantum mechanical phenomenon in which quantum 
systems exhibit correlations not possible for classical systems \cite{HHH}. 
Decoherence may be especially detrimental to highly entangled states \cite{Dur} such 
as those used for protecting quantum information. An extreme negative manifestation 
of this is ESD in which entanglement is completely lost
in finite time \cite{DH,YE1} despite the fact that the coherence loss of the system 
is asymptotic. Recently \cite{YE2}, there has been a call to develop techniques to 
counteract ESD so as to protect quantum memory from its harmful consequences.

In this paper I explore the affect of ESD on entangled states that are 
specifically examples of DFSs and NSs. My goal is to explore whether ESD 
is really a threat to quantum memories built from error aviodance schemes 
above and beyond that of typical decoherence. The direct study of the affect 
of ESD on quantum protocols has only been undertaken recently. 
In \cite{YSW1} it was shown that a three (physical) qubit error correction 
code capable of protecting a qubit of quantum information from phase flips 
is indifferent to the phenomenon of ESD. In \cite{YSW2} it was shown that in 
cluster states capable of primitive quantum gates via cluster state (or one-way) 
quantum computing protocols, correlations exist between ESD and the point at which the 
fidelity of the decohered state equals .5. In this paper such explorations are 
extended to the affect of ESD on error avoidance protocols. Other related work 
addressing ESD of multi-particle systems can be found in \cite{ACCAD,LRLSR,YYE,SB} 
and there have been several initial experimental studies of this phenomenon \cite{expt}.

As mentioned above when constructing a quantum memory it would 
be most practical to store the information and not have to address the memory
again until the information is needed. This can be achieved using error avoidance 
techniques. In addition, it is reasonable to assume that in a quantum memory the 
dominant decoherence generators would be far field, such that they affect 
all of the qubits collectively. Thus, we begin by studying a four-qubit DFS and a 
three-qubit NS that protect quantum information from collective decoherence. 
However, it is most likely that some additional decoherence that is not 
collective, but rather qubit independent, will affect the system.
This decoherence will degrade the stored quantum information and, if strong enough
may cause ESD. We will explore the relationship between the decoherence strength at which 
ESD is exhibited due to this decoherence and the fidelity of the stored qubit of 
quantum information.
 
The decoherence models we explore are the independent qubit dephasing and depolarizing environments. 
The dephasing environment is fully described by the Kraus operators
\begin{equation}
K_1 = \left(
\begin{array}{cc}
1 & 0 \\
0 & \sqrt{1-p} \\
\end{array}
\right); \;\;\;\;
K_2 = \left(
\begin{array}{cc}
0 & 0 \\
0 & \sqrt{p} \\
\end{array}
\right)
\end{equation} 
where $p$ is the dephasing strength. 
When all $n$ qubits undergo dephasing we have $2^n$ Kraus operators each of the form 
$A_l = (K_i\otimes K_j\otimes K_k)$ where $l = 1,2,...,2^n$ and $i,j,k = 1,2$. 
The depolarizing environment is described by Kraus operators
\begin{equation}
K_1 = \sqrt{1-\frac{3p}{4}}\openone,\;\; K_c = \frac{\sqrt{p}}{2}\sigma_c, \;\; c = 2,3,4,
\end{equation}
where $\sigma_c$ are the Pauli spin matrices and now $p$ is the depolarizing strength.
For the depolarizing environment there are $4^n$ Kraus operators. Though all of the below 
calculations are done with respect to $p$, I implicitly assume that $p$
increases with time, $\tau$, at a rate $\kappa$, such that 
$p = 1-e^{-\kappa\tau}$ and $p\rightarrow 1$ only at infinite times. 
I also assume equal decoherence strength on all qubits (this can
be viewed as the worst-case scenario). 

To monitor the occurrence of ESD I will utilize a number of entanglement metrics.
The first is the negativity, $N^{(i)}$, for which I will simply use the (absolute
value of the) most negative eigenvalue of the parital transpose of the system 
density matrix \cite{neg}. When there 
are more than two qubits in the system the partial transpose can be taken with 
respect to different sets of qubits $i$ giving, in general, inequivalent negativities. 
I will also make use of the two qubit concurrence \cite{conc}, $C_{jk}$. The concurrence 
between two qubits $j$ and $k$ with density matrix $\rho_{jk}$ is usually defined 
as the maximum of zero and $\Lambda$, where 
$\Lambda = \sqrt{\lambda_1}-\sqrt{\lambda_2}-\sqrt{\lambda_3}-\sqrt{\lambda_4}$
and the $\lambda_i$ are the eigenvalues of 
$\rho_{jk}(\sigma_y^j\otimes\sigma_y^k)\rho_{jk}^*(\sigma_y^j\otimes\sigma_y^k)$
in decreasing order and $\sigma_y^i$ is the $y$ Pauli matrix of qubit $i$. For the 
purposes of clearly seeing at what point ESD occurs we will use $\Lambda$ as
the concurrence noting that ESD occurs when $\Lambda = 0$ in finite time 
(i.~e.~before $p \rightarrow 1$). To measure entanglement between general 
states of three qubits I will use the tri-partite negativity \cite{SGA}, $N_3$, 
which is simply the third root of the product of the 
negativities with respect to each of the three qubits $N_3 = \sqrt[3]{N^{(1)}N^{(2)}N^{(3)}}$.

\section{Four Qubit DFS}

The smallest possible DFS that can protect one qubit of quantum information from
the affects of collective decoherence is comprised of four physical qubits \cite{ZR}. 
The two (orthogonal) basis states for this DFS are \cite{KBLW}:
\begin{eqnarray}
|0\rangle_L &=& \frac{1}{2}\left(|01\rangle-|10\rangle\right)_{1,2}\otimes\left(|01\rangle-|10\rangle\right)_{3,4}\nonumber\\
|1\rangle_L &=& \frac{1}{\sqrt{12}}(2|0011\rangle+2|1100\rangle-|0101\rangle\nonumber\\
            & & -|1010\rangle-|0110\rangle-|1001\rangle).
\end{eqnarray}
I assume an initial state of $|\psi\rangle_{DFS} = \cos a|0\rangle_L+e^{ib}\sin a|1\rangle_L$. 
Any such state will not evolve under collective decoherence 
but is degraded by independent (physical) qubit decoherence. To determine the affect of 
ESD on the storage of quantum information in this DFS we compare the decoherence strength at 
which ESD is exhibited for different entanglement measures to the fidelity of the degraded 
state. 

\begin{figure}
\includegraphics[width=4.25cm]{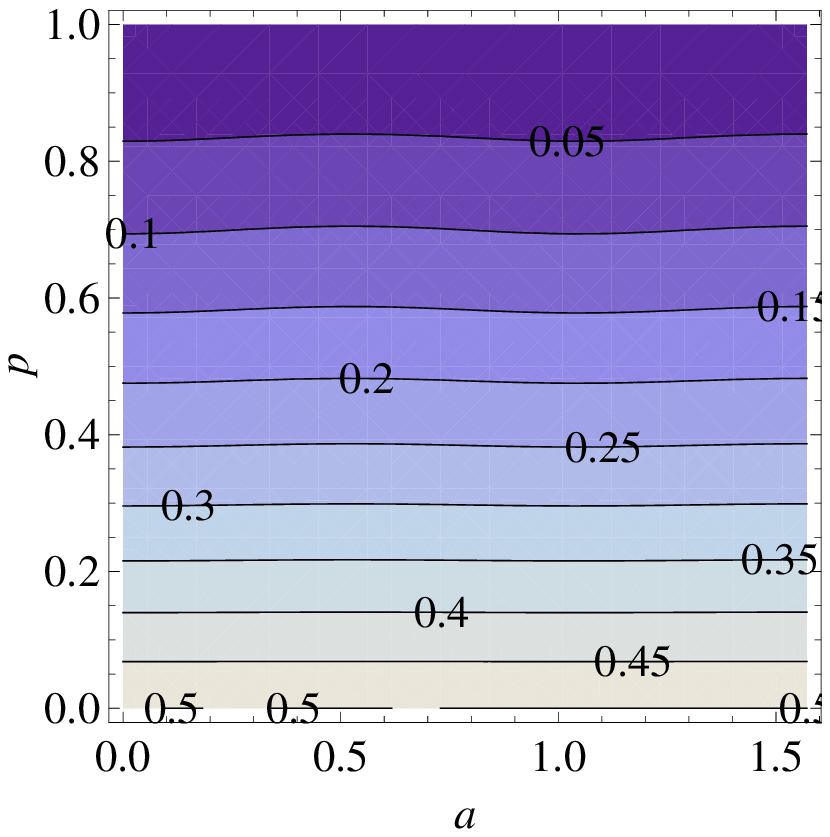}
\includegraphics[width=4.25cm]{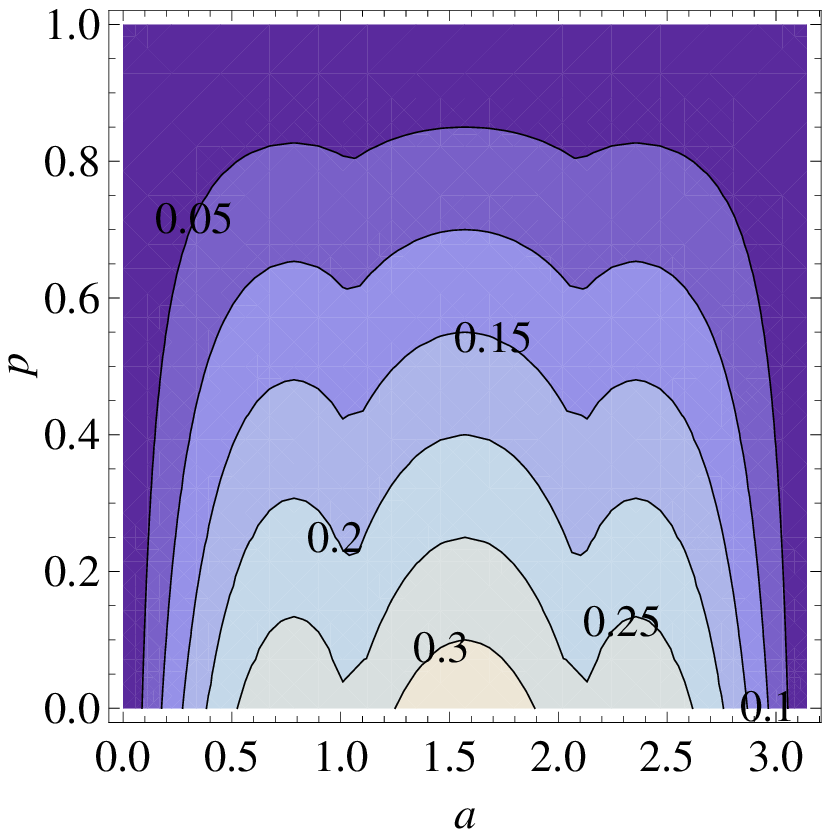}
\includegraphics[width=4.25cm]{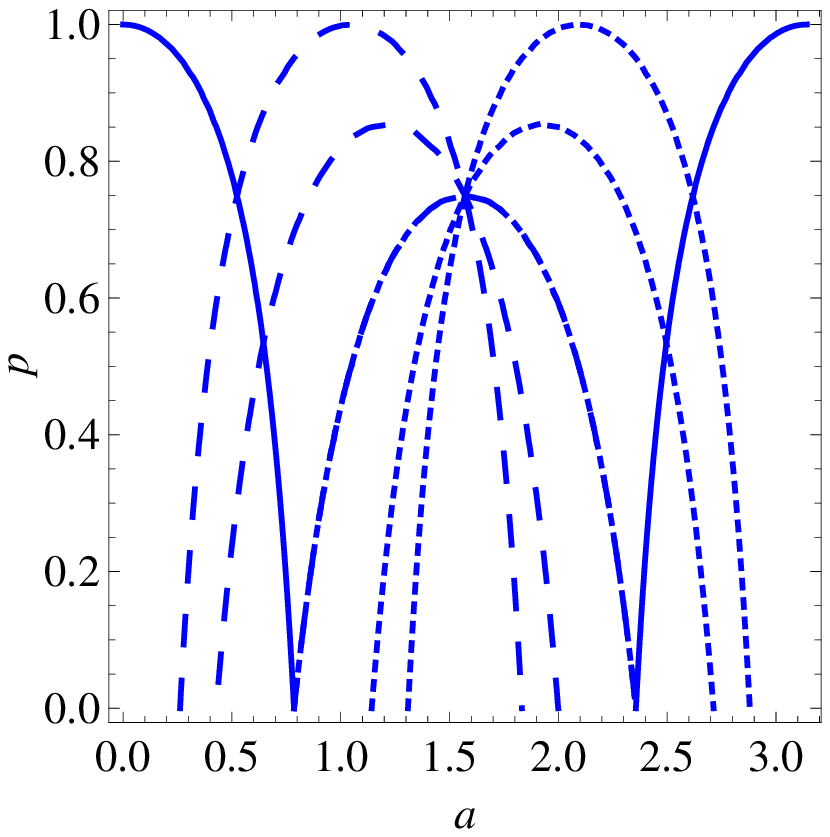}
\includegraphics[width=4.25cm]{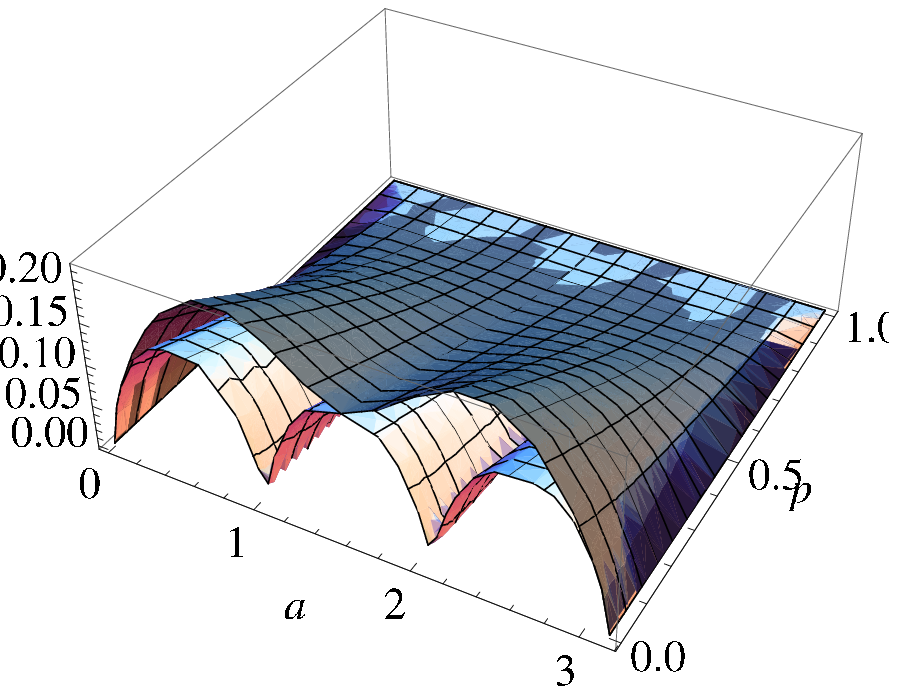}
\includegraphics[width=4.25cm]{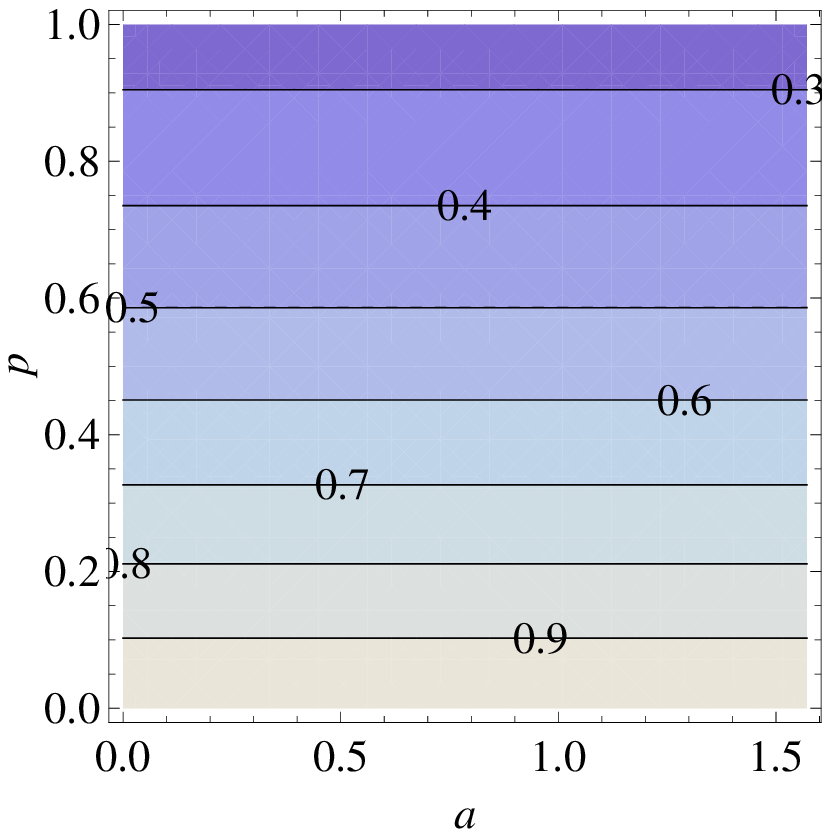}
\caption{(Color online) Negativity, $N^{(1)}$, with respect to the first qubit, (top left), 
negativity, $N^{(1,2)}$ with respect to the first two qubits (top right), curves where 
the concurrence is equal to zero (center 
left), tri-partite negativity (center right) and fidelity (bottom) of the state
$|\psi\rangle_{DFS}$, as a function of the initial state, parameterized by $a$, 
and dephasing strength $p$. For the negativity and fidelity plots $b = 0$. 
The concurrence plot shows $C_{12}$ (solid line), $C_{13}$ 
(large dashed line), and $C_{14}$ (small dashed line). ESD for $C_{12}$ is independent 
of $b$ and ESD ($C_{jk} = 0$) for the other concurrences are shown for (bottom to top) 
$b = \frac{\pi}{2}, \frac{\pi}{3},$ and 0.
The two states in the tri-partite negativity plot are 
$b = 0$ (light) and $b = \frac{\pi}{2}$ (dark). 
Note that the negativities and tri-partite negativity do not 
exhibit ESD despite the fidelity going below .5. None 
of the entanglement measures seem to be at all correlated with the fidelity of the 
state of the DFS.
}
\label{dephase}
\end{figure}

In an independent dephasing environment the fidelity of the 
four qubit system in any state $|\psi\rangle_{DFS}$ is given by:
\begin{eqnarray}
F(a,b,p) &=& \frac{1}{48}(48+p(11p-48) \nonumber\\
	   &+& p^2(\cos 4a+(2\cos 2b)(\sin 2a)^2)).
\end{eqnarray}
Yet, despite the fact that the fidelity of the state of the system can be degraded 
below .5, the system negativity does not exhibit ESD. Thus, unlike \cite{YSW2},
ESD is not an indicator that the fidelity of the stored information falls below .5.
The independent qubit dephsing environment does cause ESD for the entanglement between any two
of the qubits in the system as measured by the concurrence. For the evolution 
of concurrence between the first qubit and each of the other qubits, $C_{1j}, j = 2,3,4$, 
ESD is exhibited at different points based on the initial state as shown in Fig. \ref{dephase}. 

If concurrence in the dephasing environment does exhibit ESD while the negativity does 
not we may ask what type of entanglement is present after the concurrence goes to zero.
To answer this we can look at the tri-partite negativity, $N_3$, after tracing over one 
of the four qubits. In an independent dephasing
environment none of the tri-partite negativities exhibit ESD. Thus, the remaining entanglement
after the sudden disappearance of the concurrence between two qubits is, at least, the 
tri-partite entanglement measured by the tri-partite negativity. 

Fig.~\ref{dephase} displays
the evolution of the various entanglement metrics as a function of the intial state and 
decoherence strength and allows us to compare the onset (or not) of ESD to the evolution 
of the fidelity. There is no discontinuity or change of behavior that occurs in the fidelity 
evolution at the point where ESD sets in. In fact, there is not even a clear correlation 
between the entanglement evolution and that of the fidelity. This implies that
the affect of ESD is no more or less than that of typical decoherence. 

\begin{figure}
\includegraphics[width=4.25cm]{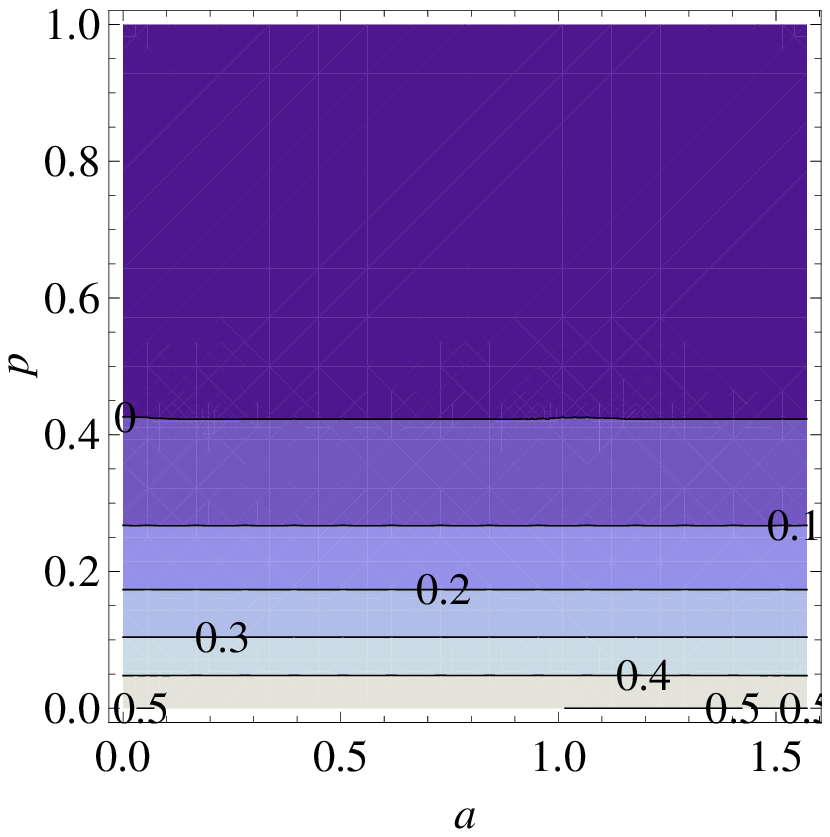}
\includegraphics[width=4.25cm]{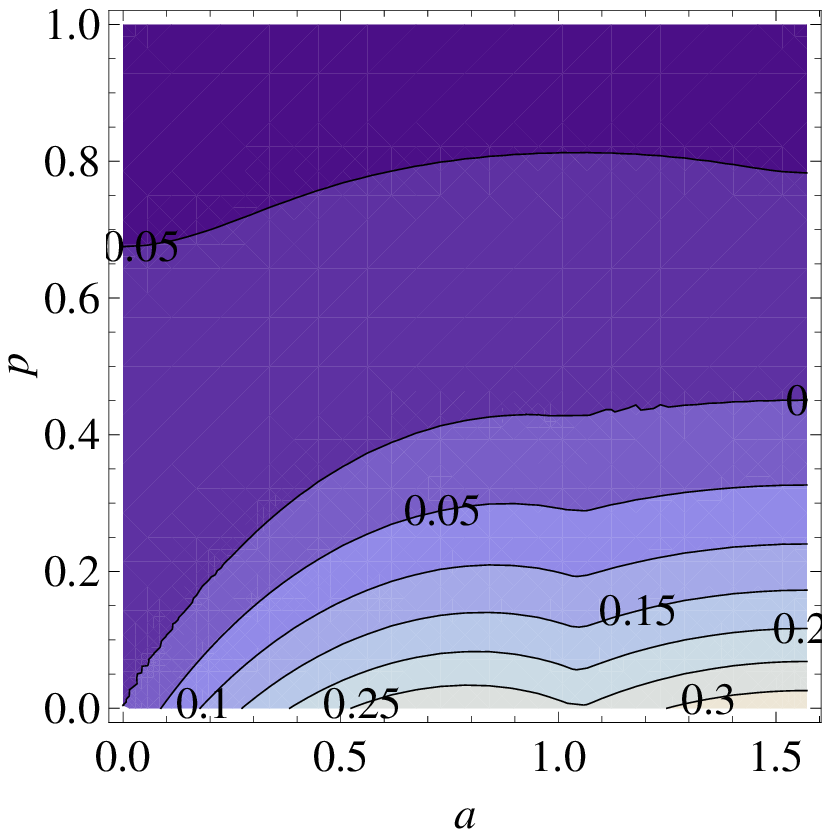}
\includegraphics[width=4.25cm]{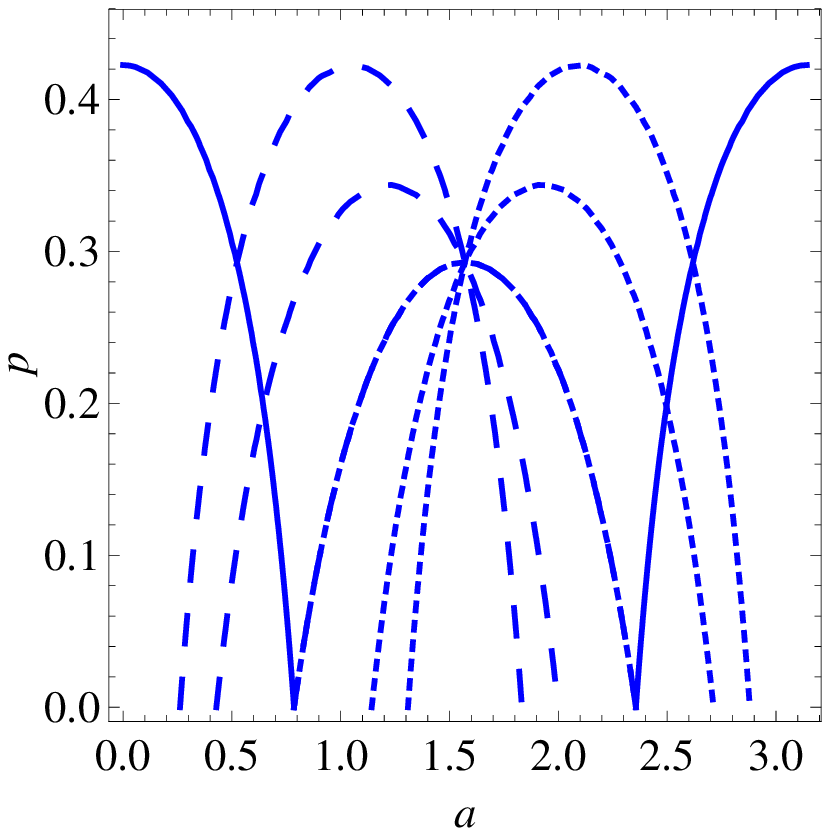}
\includegraphics[width=4.25cm]{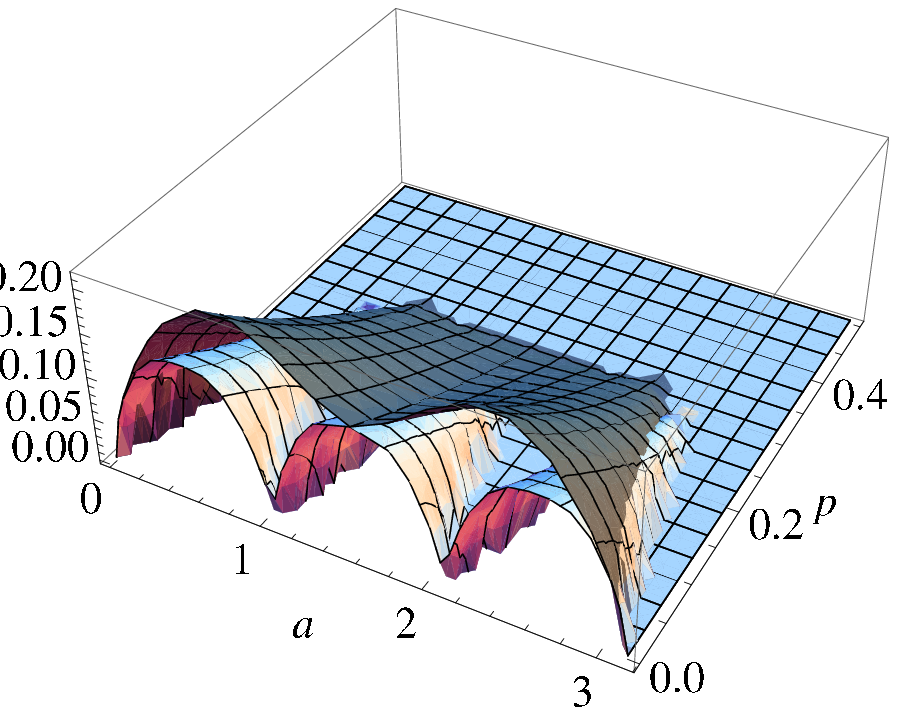}
\includegraphics[width=4.25cm]{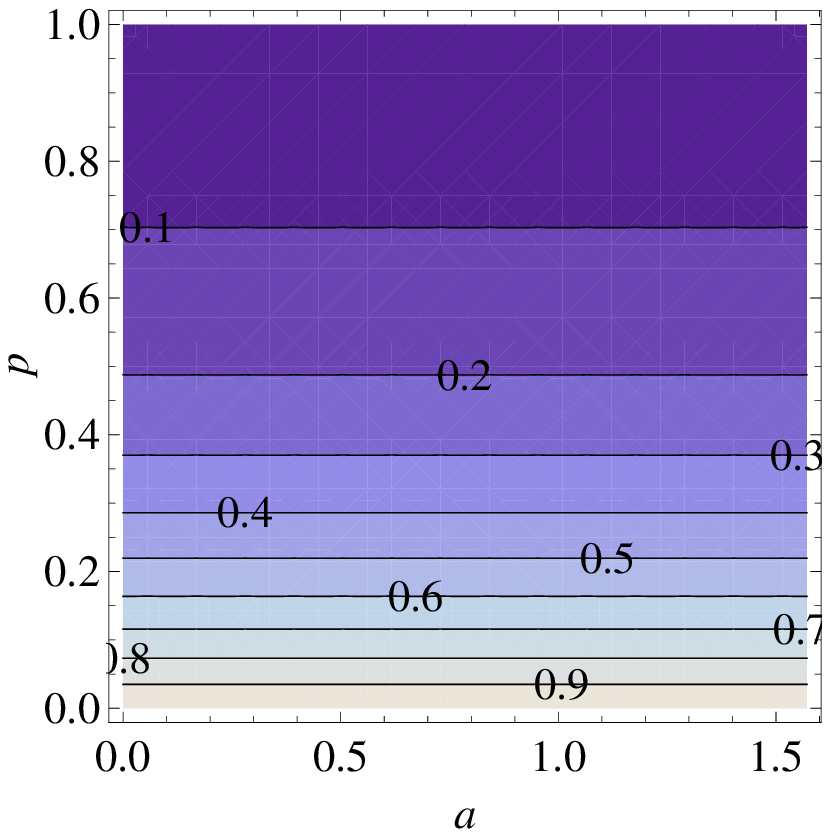}
\caption{(Color online) 
Negativity, $N^{(1)}$, with respect to the first qubit, (top left), 
negativity, $N^{(1,2)}$ with respect to the first two qubits (top right), curves where 
the concurrence is equal to zero (center 
left), tri-partite negativity (center right) and fidelity (bottom) of the state
$|\psi\rangle_{DFS}$, as a function of the initial state, parameterized by $a$, 
and depolarizing strength $p$. For the negativity and fidelity plots $b = 0$. 
The concurrence plot shows $C_{12}$ (solid line), $C_{13}$ 
(large dashed line), and $C_{14}$ (small dashed line). ESD of $C_{12}$ is independent 
of $b$ and ESD ($C_{jk} = 0$) for the other concurrences are shown for (bottom to top) 
$b = \frac{\pi}{2}, \frac{\pi}{3},$ and 0.
The two states in the tri-partite negativity plot are 
$b = 0$ (light) and $b = \frac{\pi}{2}$ (dark). 
All of the entanglement metrics now exhibit ESD but there is no correlation 
between when this occurs and the fidelity of the state of the DFS. 
}
\label{depol}
\end{figure}

In an independent qubit depolarizing environment the fidelity of the 
four qubit system in any state $|\psi\rangle_{DFS}$ is given by:
\begin{eqnarray}
F(a,b,p) &=& \frac{1}{16}(p^2(p-1)^2(\cos 4a+\cos 2b(1-\cos 4a))\nonumber\\
        &+& 8p^4-34p^3+59p^2-48p+16).
\end{eqnarray}
However, in this case the negativities with the partial transpose taken with respect to one, $N^{(j)}$,  
or two qubits, $N^{(jk)}$, do exhibit ESD. The two qubit concurrence (taken between the same qubit 
combinations as studied in the dephasing environment) also exhibits ESD, though at lower decoherence 
strengths than in a dephasing environment, as does the the tri-partite 
negativity. The evolution of these entanglement metrics is shown in Fig.~\ref{depol}.

In asking what type of entanglement lasts the longest we note that 
in the plots below, the negativity metrics disappear at $p$ slightly greater than 0.4. The concurrence
between any of the two-qubit combinations usually disappears at much weaker decoherence 
strengths while the tri-partite negativity disappears at $p < 0.4$. Thus, the type of entanglement 
that drives the negativity to not exhibit ESD until $p > 0.4$ is some sort of three qubit 
entanglement not measured by the tri-partite entanglement or genuine four-partite entanglement. 

The behavior of the negativity $N^{(j)}$ is similar to that of the fidelity in that there is little 
dependence on $a$ or $b$. $N^{(1)}$ for initial state $a = b = 0$ exhibits ESD at 
$p \simeq .4227$ and at that value the fidelity is about $.6220$. However, this should be 
compared to the dephasing environment where the fidelity can fall below .5 while no ESD is exhibited.
This comparison highlights the lack of correlation between ESD and the proper functioning of a quantum 
memory. In addition, both $N^{(j)}$ and the fidelity  
differ from the behavior of the negativity where the partial transpose is taken with respect 
to two qubits. Finally, neither the behavior of the concurrence nor of the tri-partite 
negativity are at all correlated with the behavior of the fidelity. The most we can say is 
that as depolarization strength decreases so does the fidelity and the amount of entanglement 
but there is no correlation between these parameters.

\section{Three Qubit NS}

While a DFS requires four physical qubits to protect against collective decoherence, a noiseless
subsystem (NS) can provide the same protection using only three physical qubits. This is done 
by storing the quantum information in the total angular momentum $S = 1/2$ subspace and storing
the quantum information in the two pathways leading to the total angular momentum. 

Storing the information in this way, and not in the state of the system, means that the quantum 
information is protected and can be efficiently extracted despite the fact that the actual state
of the system is affected by the decoherence. For the three qubit NS the states which span each 
of the two logical qubit basis states are:
\begin{eqnarray}
|0\rangle_L\otimes|+1/2\rangle_Z &=& \frac{1}{\sqrt{3}}(|001\rangle+\omega|010\rangle+\omega^2|100\rangle)\nonumber\\
|0\rangle_L\otimes|-1/2\rangle_Z &=& \frac{1}{\sqrt{3}}(|110\rangle+\omega|101\rangle+\omega^2|011\rangle)\nonumber\\
|1\rangle_L\otimes|+1/2\rangle_Z &=& \frac{1}{\sqrt{3}}(|001\rangle+\omega^2|010\rangle+\omega|100\rangle)\nonumber\\
|1\rangle_L\otimes|-1/2\rangle_Z &=& \frac{1}{\sqrt{3}}(|110\rangle+\omega^2|101\rangle+\omega|011\rangle),
\end{eqnarray}
where $L$ refers to the logical qubit which is protected against collective
errors, $Z$ is the subsystem that experiences the errors, and $\omega = e^{2\pi i/3}$.
Following the protocols of \cite{VF,FV}, we encode the intial single qubit state 
$\cos a|0\rangle+e^{ib}\sin a|1\rangle$ into the state 
$(\cos a|0\rangle_L+e^{ib}\sin a|1\rangle_L)\otimes|-1/2\rangle_Z$. To calculate 
the fidelity of the stored information after decoherent evolution we apply the decoding 
circuit of \cite{VF,FV} and compare the single qubit output to the initial state (which 
is equivalent to encoding and decoding with no applied decoherence).

All of the above states spanning the $S = 1/2$ subspace contain some entanglement. Does the 
finite time loss of this entanglement affect the system's ability to protect the stored quantum information?
As mentioned above, a qubit of information stored in this NS is perfectly protected against 
collective decoherence despite the fact that the state of the system after application of the 
decoherence is not equivalent to the initial encoded state. In fact, the fidelity between the 
intial state and the state subject to collective decoherence may fall as low as .5. However, the 
fidelity of the stored quantum information is 1 and the state of the system does not
exhibit ESD. 

Qubit-independent decoherence can cause ESD of the three qubit NS system. To see how this affects the 
stored information we compare the decoherence strengths at which ESD occurs to the fidelity of the 
stored qubit of information.
In an independent dephasing environment with the dephasing strength on all three qubits equal 
to $p$, the system does not exhibit ESD neither with respect to any of the system negativity 
measures nor with respect to the concurrence between any two of the three qubits. 
The fidelity of the stored quantum information, given by 
\begin{equation}
F(a,p) = \frac{1}{12}(12 - 5p - p(2\cos 2a + \cos 4a)),
\end{equation}
is not dependent on $b$, and can go as low as $\frac{1}{3}$.   

\begin{figure}
\includegraphics[width=4.25cm]{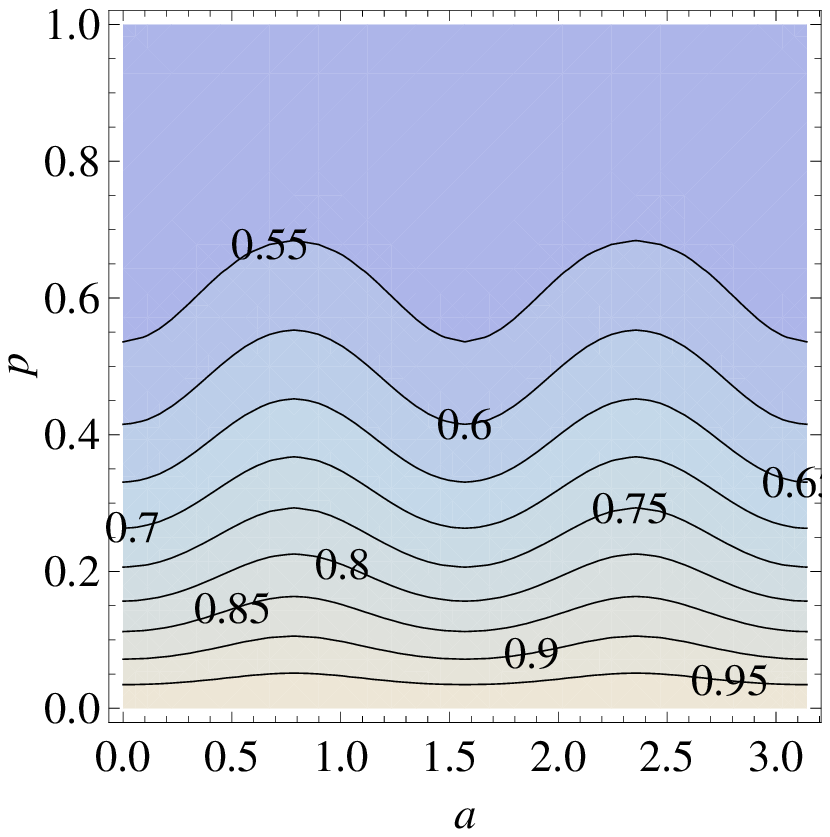}
\includegraphics[width=4.25cm]{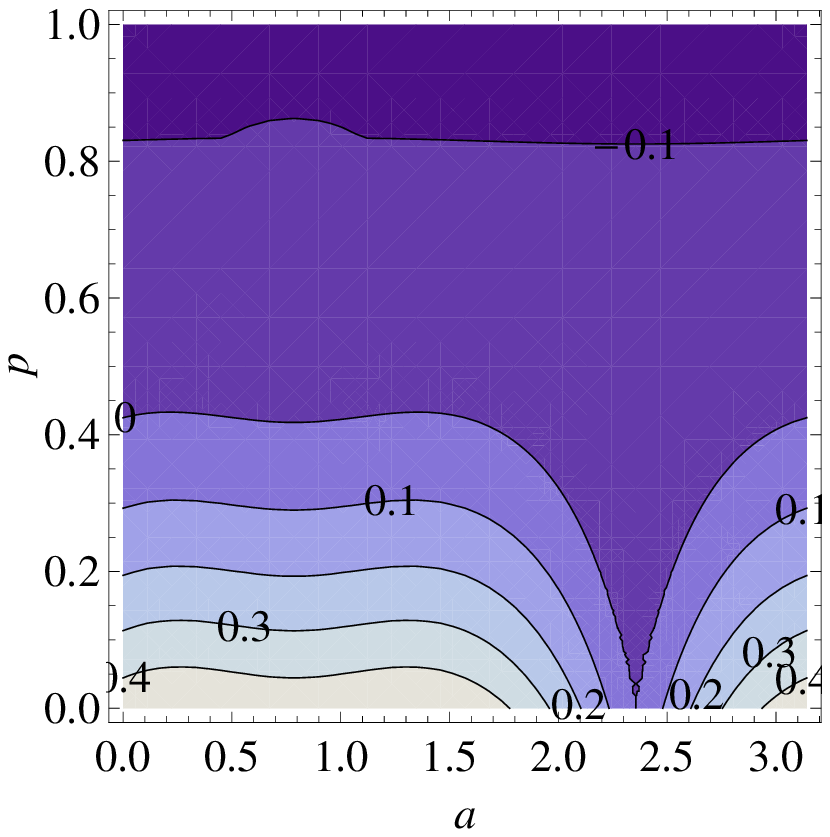}
\includegraphics[width=4.25cm]{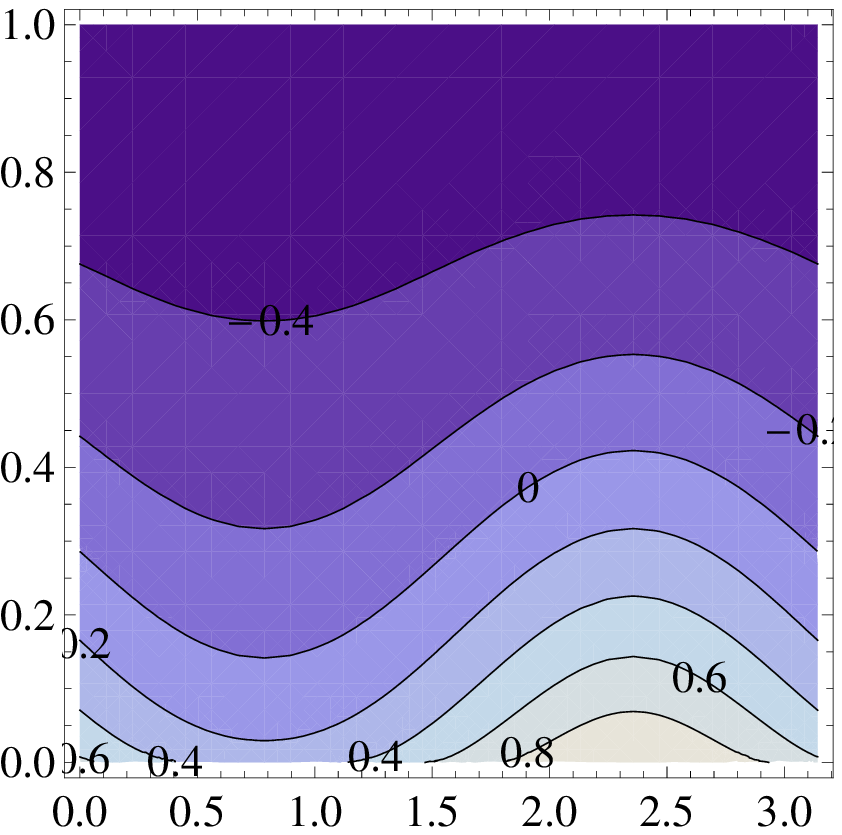}
\includegraphics[width=4.25cm]{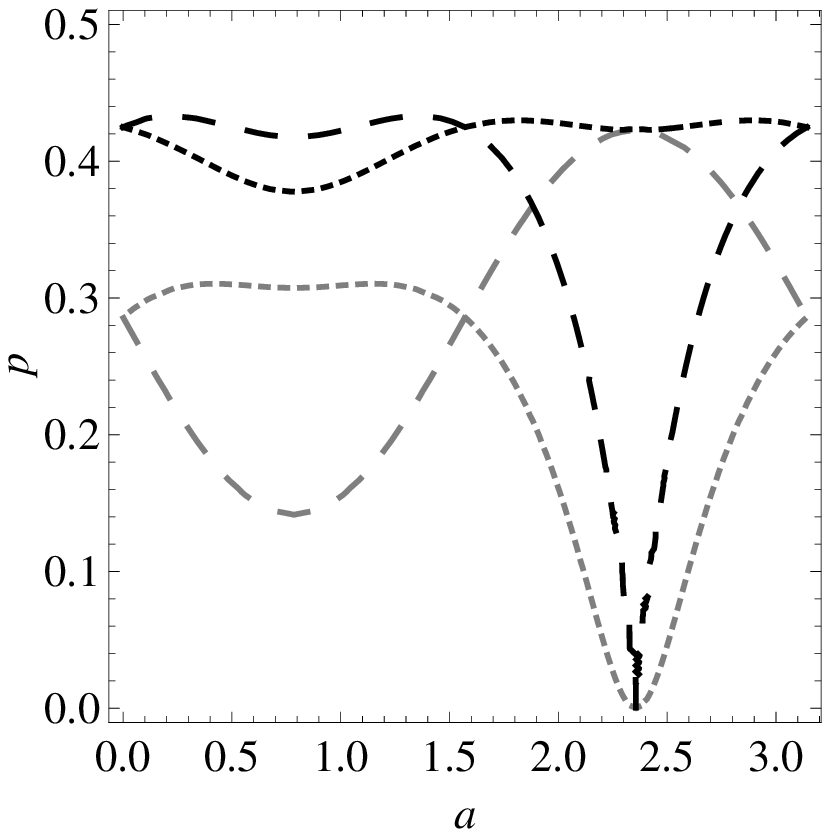}
\caption{(Color online) Top: Fidelity (left) and negativity with 
repsect to the third qubit (right) for the three qubit NS as a function of 
decoherence strength and initial state paramaterized by $a$ (for all entanglement measures 
the figure shows the case where the intial state has $b = 0$). Bottom: Concurrence between qubits
1 and 2 (left) and ESD of concurrence versus ESD of negativity (right): light lines show
where ESD is exhibited by the concurrence between qubits 1 and 2 (dashes) and 1 or 2 and 
3 (dots), dark lines show where ESD is exhibited by the negativity with the partial 
transpose taken with respect to the third qubit (dashes) and first or second qubit (dots).
ESD for the different entanglement metrics is exhibited at different depolarization strengths. 
The difference between these strengths implies the survival of tri-partite entanglement
after the decay of bi-partite entanglement. There does not appear to be any sort of 
correlation between the onset of ESD for any of these metrics and the fidelity of the 
stored quantum information.}
\label{NSfig}
\end{figure}

In a qubit-independent depolarizing environment, with depolarization strength $p$ on each
qubit, the system does exhibit ESD for a host of entanglement measures. We compare the 
decoherence value at the onset of ESD to the fidelity of the stored quantum information 
which is given by 
\begin{equation}
F(a,p) = \frac{1}{4}(4-p(5+p(p-4))-p(p-1)^2\cos 4a).
\end{equation}
Note that the fidelity is again not dependent on $b$, and at $p = 1$ the fidelity goes to $1/2$. 
ESD of the negativity occurs at similar decoherence values irrespective of the choice of 
partial transpose. The fidelity of the stored quantum information and the negativity with
partial transpose taken with respect to the first qubit are shown in Fig.~\ref{NSfig}. Note 
that for $p \simeq .42486$ ESD is observed for states where $a = 0,\frac{\pi}{2}$. The fidelity
where ESD is exhibited for those states is $\simeq .595$. Compare this to the case of dephasing where 
no ESD is exhibited and the fidelity can be as low as $\frac{1}{3}$ \cite{footnote}.

Under independent qubit depolarization the concurrence between any of the two qubits of the system
also exhibits ESD as shown in Fig.~\ref{NSfig}. In general ESD of concurrence occurs at lower 
decoherence strengths than ESD of the negativity implying remaining tri-partite entanglement 
in the system after the disappearance of the bi-partite entanglement. Again no correlation is
seen between the decoherence strength where ESD is exhibited and the fidelity.

\section{Other Protected Subsystems}

There are some DFS and NS variants which do not exhibit ESD, or even utilize
any entanglement at all. For example, the states $|0\rangle_L = |01\rangle$,
$|1\rangle_L = |10\rangle$, form a two qubit DFS to protect against 
collective dephasing \cite{FVHTC}. General states within this logical 
basis have EPR-type entanglement which do not exhibit ESD in a 
depolarizing environment (the two qubit density matrix does not 
have the $\times$ form \cite{YE1}).  

The parity of two qubits forms an NS that can protect against collective 
bit flip errors \cite{FV}. Initial states within this space are not entangled
at all. While collective $\sigma_x$ rotations not of $\pi/2$ can cause
entanglement, such entanglement is again not subject to ESD. 

For completeness, I have looked at a couple of additional examples of DFSs
where the system does exhibit ESD 
to see if any correlation can be found between the finite-time loss of entanglement 
and the fidelity of stored quantum information. Specifically, I looked at DFSs
which consist of a doubly degenerate ground state of a three \cite{DFS1} and four 
\cite{DFS2} qubit system with always-on Heisenberg couplings. The energy gap
between the logical qubit states and other states of the system forces decoherence 
generators to add energy in order to affect the system state.  

In a series of calculations comparing ESD of different types of entanglement with 
the fidelity of the state of the DFS, no correlation is found. This again shows that
ESD does not affect the workings of a quantum memory.

\section{Conclusion}

In conclusion, I have explored several qubit systems that one would typically
utilize as a quantum memory. I have shown that the disappearance of entanglement
from these systems does not correlate with the loss of fidelity of 
the stored quantum information. Certainly there is no change in the behavior of
the fidelity when the system undergoes ESD, and there is not even a correlation
between the decoherence strengths where ESD is exhibited and a given fidelity
measure. The systems I explored were a four qubit decoherence free subspace
and a three qubit noiseless subsystem. Both of these systems can protect a qubit of quantum 
information from collective decoherence but cannot protect quantum information 
from independent qubit decoherence which can thus cause ESD.

It is always necessary to protect quantum information from the possibly debilitating
affects of decoherence. ESD may be caused by certain decoherence generators but
protection of quantum information from the ESD phenomenon does not require any special 
attention.   

It is a pleasure to thank L. Viola and G. Gilbert for helpful feedback 
and acknowledge support from the MITRE Technology Program under MTP grant \#07MSR205.



\begin{thebibliography}{99}
\bibitem{footnote}
It is interesting that the lowest possible fidelity for the independent 
depolarizing environment is higher than the lowest possible fidelity for 
the independent dephasing environment.
\end{thebibliography}
\end{document}